\documentclass{article}
\usepackage{spconf,amsmath,graphicx}
\usepackage{multirow}
\usepackage{subfigure}
\usepackage{float}
\usepackage{color}
\usepackage{stfloats}
\usepackage{booktabs}
\usepackage{cuted}

\usepackage{enumitem}

\title{FILLER WORD DETECTION WITH HARD CATEGORY MINING AND \\ INTER-CATEGORY FOCAL LOSS}

\name{Zhiyuan Zhao$^{1}$, Lijun Wu$^{1}$, Chuanxin Tang$^{1}$, Dacheng Yin$^{2}$, Yucheng Zhao$^{2}$, Chong Luo$^{1}$}
\address{
$^{1}$Microsoft Research Asia, Beijing, China \\
$^{2}$University of Science and Technology of China, Hefei, China
}
\begin{document}
%

\maketitle
\begin{abstract}
Filler words like ``um" or ``uh" are common in spontaneous speech. It is desirable to automatically detect and remove them in recordings, as they affect the fluency, confidence, and professionalism of speech.
Previous studies and our preliminary experiments reveal that the biggest challenge in filler word detection is that fillers can be easily confused with other hard categories like ``a" or ``I". 
In this paper, we propose a novel filler word detection method that effectively
addresses this challenge by adding auxiliary categories dynamically and applying an additional inter-category focal loss.
The auxiliary categories force the model to explicitly model the confusing words by mining hard categories. 
In addition, inter-category focal loss adaptively adjusts the penalty weight between ``filler" and ``non-filler" categories to deal with other confusing words left in the ``non-filler" category. Our system achieves the best results, with a huge improvement compared to other methods on the PodcastFillers dataset.

\end{abstract}
\begin{figure*}[t]
  \centering
  \includegraphics[width=\linewidth]{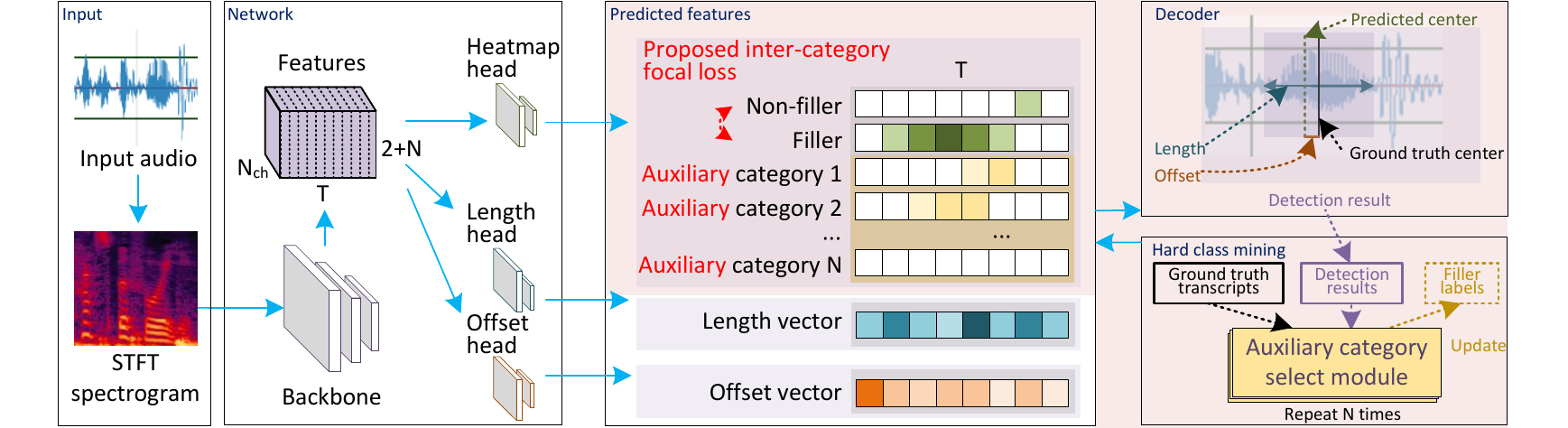}
  \caption{Overview of our proposed method. Hard category mining module can select auxiliary categories. Inter-category focal loss is applied on ``Non-filler" and ``Filler" categories.}
  \label{fig:pipeline}
\end{figure*}

\section{Introduction}
Those who do not have professional training in speaking tend to use filler words such as ``um" or ``uh" frequently \cite{filler-adobe,filler-increase, overview-disf-congnitive}. 
Filler words make speech sound unprofessional, reducing the speaker's credibility \cite{overview-exploring, overview-disf-congnitive, overview-invest-fluency}. In multimedia creation scenarios, e.g., podcasts, voice conferences, voice interviews, voice messages, or video demos, content creators desperately need an automatic tool to detect and remove filler words to save their time and effort in speech editing. There are many works devoted to filler word detection task.

Several Automatic Speech Recognition (ASR) based filler word detection methods have been investigated \cite{asr-clinician, asr-semi-markov, asr-segmentation, asr-reviewer}. In these systems, speech is first recognized by an ASR model, and then 
filler words are detected in the transcript.
But it is challenging for most ASR systems to 
transcribe filler words since the systems are often trained on spoken text corpora that contain almost no filler words \cite{filler-adobe}. Besides, some studies have proven that even after re-training an ad-hoc ASR system on a dataset with filler words, it is still difficult to achieve better filler word detection performance than a specialized filler word detection model \cite{asr-e2e}, not to mention the huge computational cost of training and using an ASR model. 
AVC-FillerNet \cite{filler-adobe} also leverage the ASR model, but they use ASR to discard regions with transcribed words to get filler candidate regions.

Some ASR-free based
disfluencies, social signal, or stuttering detection methods have been proposed \cite{filler-social, disf-prosody, stutter-fluentnet}. They can also detect filler words or interjections, which is another term for filler words.
In addition to filler word detection, they can also detect other sound events such as laughter, long silences, repetition, or correction, etc. These methods need to detect multiple events at the same time, and they require data with more comprehensive labels, which makes their design complex.

More recently, some ASR-free methods that mainly focus on filler words have been proposed \cite{filler-increase, filler-adobe}. Sagnik et al. proposed a system that can detect filler words and remove them from speech \cite{filler-increase}. They use a Convolutional Recurrent Neural Network (CRNN) for filler word detection. However, they have observed that the system makes it difficult to distinguish speech segments that sound similar to ``uh" or ``um". VC-FillerNet \cite{filler-adobe} is a two-stage ASR-free filler detection method. 
It first leverages a Voice Activity Detection (VAD) module to find voice regions, and then uses a classification model to predict frame-level classification results.

In the conventional methods, all non-filler words are mixed up to form an ``unknown" or an ``others" category. Based on the observation from previous work \cite{filler-increase} and our preliminary experiment (Section~\ref{sec:preliminary}) that filler words are easily confused with non-filler words that sound similar to filler. 
In this paper, we proposed a novel end-to-end ASR-free filler word detection method with dynamic auxiliary category selection and an additional inter-category focal loss.  
We add some auxiliary categories based on top false positive (confusing) words as a hard category mining strategy, to force the network to explicitly model the confusing words. 
With our new training scheme, we can select auxiliary categories automatically with just one training loop.
In addition,
we designed an inter-category focal loss which can adaptively control the penalty weight between any two categories. In our setting, we applied the inter-category focal loss on the ``non-filler" category and the ``filler" category to deal with other hard categories left in the ``non-filler" category besides the auxiliary categories we selected. Our system achieves State-Of-The-Art (SOTA) results on the PodcastFillers dataset \cite{filler-adobe}. The ablation experiments prove that the auxiliary categories and the inter-category loss we designed are effective.

\section{METHOD}

\subsection{Preliminary experiment (challenge of the problem)}
\label{sec:preliminary}
\begin{figure}[t]
  \centering
  \includegraphics[width=\linewidth]{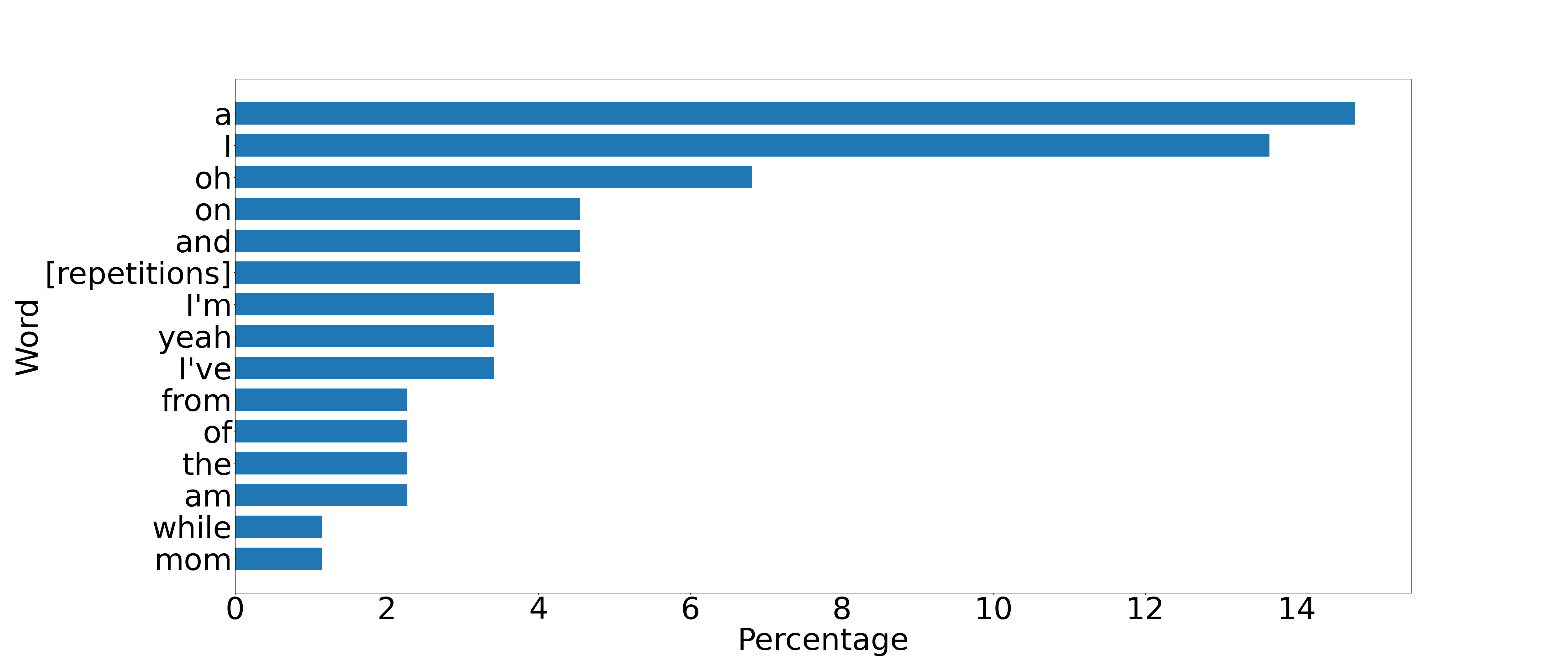}
  \caption{Top false positive (FP) words from our preliminary results on the PodcastFillers validation dataset.}
  \label{fig:fp}
\end{figure}

Filler words only contain one syllable. There are many single words or parts of multi-syllable words that have similar pronunciations. Sagnik et al. \cite{filler-increase} have observed that their system 
has difficulty in distinguishing speech segments that sound similar to ``uh" or ``um", such as ``a", ``the" or the beginning part of ``umbrella". We did a preliminary experiment to study this phenomena by calculating the confusion matrix of the classification results. Fig.~\ref{fig:fp} shows the top false positive (FP) analysis from our preliminary results on the PodcastFillers validation dataset. In this paper, we use ``confusing words (categories)" indicate words (categories) that are easily mis-classified. We found that confusing words are largely concentrated on certain words. 
Noting that in conventional methods (we discussed above), these confusing words are simply mixed with other words to form an ``unknown" or ``others" category, hence in this preliminary study, we have used focal loss to help with this problem for hard sample mining. Even though, according to the analysis results shown in Fig. \ref{fig:fp}, there are still many mis-classifications left over. This strongly motivates us to design better approach and algorithm to address this problem. 


\subsection{Overview of our method}
Shortly speaking, we design a new scheme to dynamically select auxiliary categories during training, and a new inter-category focal loss to apply on the specified categories.
Since our goal is to detect specific short voice signals in continuous speech, the task can be viewed as a Keyword Spotting (KWS) when the keywords are filler words. Therefore, we adapt AF-KWS \cite{baseline-afkws} to build our backbone model. As shown in Fig. \ref{fig:pipeline}, our method is an end-to-end ASR-free method. It takes raw speech as input, extracts Short-Time Fourier Transform (STFT) features, then uses a Residual Network (ResNet) \cite{baseline-resnet} to extract high-level features, and then employs three trainable prediction heads to regress keyword heatmap, word length, and offset, respectively. It can directly output the location of detected keywords after a decoder module. Our training scheme can automatically select confusing words as a new category and then continue training. Now we introduce the details of our proposed two mechanisms in the follows.


\subsection{Auxiliary category selection}
Since the ``filler words" and ``non-filler words" are largely imbalanced (lots of non-filler words and hard confusing words), we adopt focal loss~\cite{loss-focalloss} to alleviate training difficulty, which can dynamically increases the penalty for hard samples.
However, these non-filler words are simply mixed in one ``non-filler" category, which is still hard for the focal loss to learn the features of the confusing (hard) words. 


To reduce the mis-classification of confusing words and filler words, we add some confusing words as separate auxiliary categories to force the model to explicitly learn independent features from these confusing words.
However, the detected confusing words may be different in different training stages for each dataset, and we need to pick some of the confusing words as new categories after model training, then adjust the model classification layer and re-train the model.
Obviously, naively repeating this process is very complicated and time-consuming. For this reason, we design a training scheme to dynamically add confusing words as auxiliary categories one by one so that the model can distinguish confusing words through only one-pass training.

As shown on the right part of Fig. \ref{fig:pipeline}. We first add $h$ classification heads as placeholders. As the training continues, every $k$ iterations, we perform evaluation and analyze the most confusing word in the current training state, adding it to a placeholder as a new category for subsequent training, repeating this process until all the placeholders are filled.


\subsection{Inter-categories focal loss}

We have picked the most confusing words into separate categories. However, since the capacity of the model backbone is limited, we cannot add too many additional auxiliary categories and there still remain some confusing words in the ``non-filler" category.

Focal loss \cite{loss-centernet, loss-focalloss} focuses training on a sparse set of hard examples and prevents the vast number of easy negatives from overwhelming the detector during training. 
It will treat hard samples in all categories indiscriminately, but the difficulty of distinguishing is different for hard samples in the auxiliary categories that we have picked and those that still exist in non-filled categories.

We design an inter-category focal loss to address the problem. It can focus on any two categories, allows one to control the weight of the loss of positive and negative samples, and can adjust the penalty adaptively.
The form of inter-category focal loss is: 
\begin{equation}
{L_{_{AB}}} =  - \frac{1}{N}\sum\limits_{t,c = {c_A}} {{{\left( {{{\hat Y}_{t,{c_B}}}} \right)}^\gamma }\left( {{\mu _{AB}}{L_{pos}} + {\omega _{AB}}{L_{neg}}} \right)} 
\end{equation}
\begin{equation}
{L_{pos}} = {\left( {1 - {{\hat Y}_{t,{c_A}}}} \right)^\alpha }\log \left( {{{\hat Y}_{t,{c_A}}}} \right)
\end{equation}
\begin{equation}
{L_{neg}} = {\left( {{{\hat Y}_{t,{c_A}}}} \right)^\alpha }\log \left( {1 - {{\hat Y}_{t,{c_A}}}} \right),
\end{equation}
where $ {L_{_{AB}}} $ focus on the samples where the model mis-classified class $ A $ into $ B $. Where $ N $ is the sample number, $t$ is the point in the time dimension, ${c = {c_A}} $ means we focus on the samples in class $ A $, 
${L_{pos}} $ is positive loss for class $ A $ while 
${L_{neg}} $ is negative loss, 
$ {{{({{{\hat Y}_{t,{c_B}}}})}^\gamma }} $
is dynamic weight, which comes from the probability that class $ A $ is wrongly classified into $ B $, $ \alpha $ and $ \gamma $ are weight factors for $ {L_{_{AB}}} $, $ \mu $ and $ \omega $ control the balance between positive and negative losses.


The overall training loss consists of an optimized focal loss \cite{loss-centernet}
\begin{small}
\begin{equation}
\label{equ:focal loss}
{L_{main}} = - \frac{{1}}{N}\left\{ {\begin{array}{*{20}{l}}
  {{\text{   }}\sum\limits_{t,c} {{{\left( {1 - {{\hat Y}_{t,c}}} \right)}^\alpha }\log \left( {{{\hat Y}_{t,c}}} \right)} {\text{ if }}{Y_{t,c}} = 1,} \\ 
  {\begin{array}{*{20}{l}}
  {\sum\limits_{t,c} {{{\left( {1 - {Y_{t,c}}} \right)}^\beta }{{\left( {{{\hat Y}_{t,c}}} \right)}^\alpha }} } \times \\ 
  {\sum\limits_{t,c} {\log \left( {1 - {{\hat Y}_{_{t,c}}}} \right)} } 
\end{array}{\,\quad\text{    otherwise}}} 
\end{array}} \right.
\end{equation}
\end{small}
and a pair of inter-category focal losses between the ``non-filler" and the ``filler" categories:
\begin{equation}
\label{equ:overall}
L = {L_{main}} + {L_{_{FN}}} + {L_{_{NF}}},
\end{equation}
where $L_{FN}$ means the loss that model misclassified ``Filler" into ``Non-filler", defined by $L_{AB}$, and $L_{NF}$ is defined in a same way.

\subsection{Implementation details}

We train our model with a learning rate initialized at 5e-3. The learning rate will be reduced to one-tenth of the previous one at the 300th, 350th, and 400th epoch. We train a total of 450 epochs. For factors in focal loss, we set $ \alpha=2 $, $ \beta=4 $. For data preprocessing, data augmentation, feature extraction, and model backbone settings, we follow the settings of the AF-KWS \cite{baseline-afkws} method. For dynamically adding auxiliary heads, we add auxiliary heads starting at the 120th epoch and add one auxiliary head every $k=$10 epochs. 

\section{EXPERIMENTS}

\subsection{Experimental settings}

We evaluate the proposed method on the PodcastFillers datasets~\cite{filler-adobe}. 
We compare our system with three strong baselines: a two stage method VC-FillerNet \cite{filler-adobe}, a neural-network based method Filler-CRNN \cite{filler-increase} and a forced-aligner based method, Gentle \cite{baseline-gentle}. Besides VC-FillerNet, Ge et al. \cite{filler-adobe} proposed an ASR-based method, AVC-FillerNet, which leverages results from a trained ASR model in their work. We exclude this method since other baseline methods and our method are ASR-free.

We compute event-based metrics (Precision, Recall, F1) \cite{loss-metrics} for the ``filler" category and follow \cite{filler-adobe} to choose a 200 ms tolerance level.
We exclude segment-based metrics in the work since our method can detect filler words from continuous speech directly and does not need to split speech into short segments for classification. 

\subsection{Results and comparison}

\begin{table}[t!]

  \centering
  \begin{tabular}{lccc}
    \toprule
    \multicolumn{1}{c}{\textbf{Method}} & 
    \multicolumn{1}{c}{\textbf{Precision}} & 
    \multicolumn{1}{c}{\textbf{Recall}} & 
    \multicolumn{1}{c}{\textbf{F1}} \\
    \midrule
   Gentle* & 	        77.0& 		   64.9&			70.4  \\
   Filler-CRNN* & 		37.5& 		   78.3& 			50.7  \\
   VC-FillerNet & 		66.0& 		   76.9& 			71.0  \\
   Ours & 		        80.9& 		   81.0& 			81.0\\
    \bottomrule
  \end{tabular}
  \caption{Performance comparison on the PodcastFillers dataset. The methods with an * mean a re-implemented version of the PodcastFillers dataset by \cite{filler-adobe}.}
  \label{tab:overview-podcast}
\end{table}


The results are shown in Table \ref{tab:overview-podcast}. 
We see that the proposed method significantly outperforms all the other systems. When the number of auxiliary categories $ h=8 $, the auxiliary categories obtained by the model are: ``and", ``a", ``[repetitions]", ``[breath]", ``the", ``I", ``[agree]" and ``oh". Apparently, they are pronounced similar to filler words and are easily confused.


\begin{figure}[htbp]
    \centering
    \subfigure[Our method without auxiliary categories and inter-category focal loss]{

        \centering
        \includegraphics[width=0.47\linewidth]{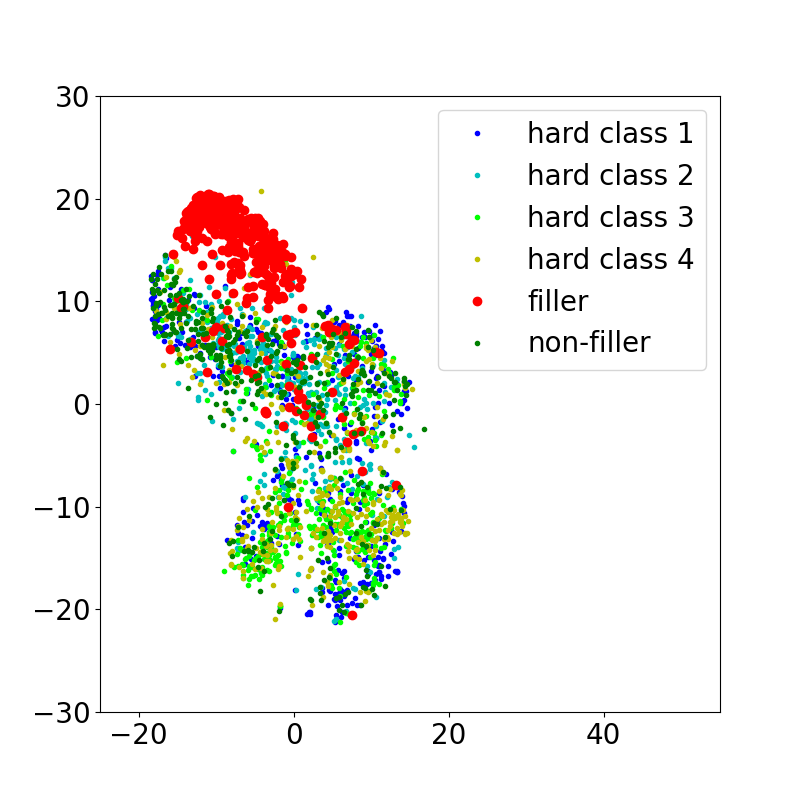}

        \label{tsne-base}
    }
    \subfigure[Our method with auxiliary categories and inter-category focal loss]{
	    \centering
	    \includegraphics[width=0.47\linewidth]{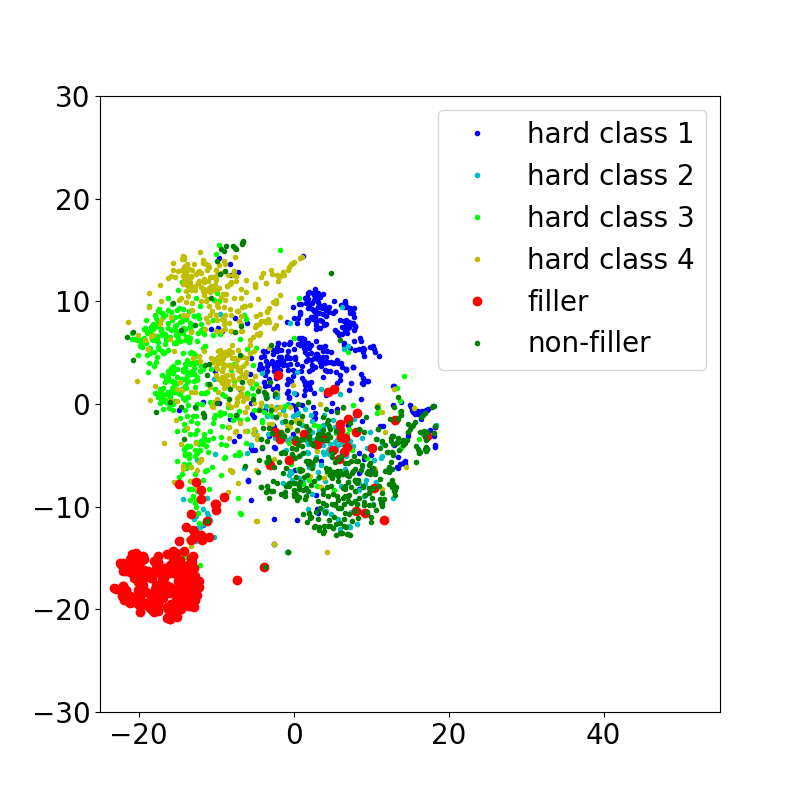}
        \label{tsne-ours}
    }
    \caption{Visualization of the latent feature embeddings learned by our model using t-SNE projection.}
    \label{fig:inter- and intro-class distance}

\end{figure}

Fig. \ref{fig:inter- and intro-class distance} shows a visualization of the latent feature embeddings learned by our model using t-SNE \cite{t-sne}. Hard classes 1-4 are selected by our method from the non-filler category. Our proposed method presents good discrimination of filler words and non-filler words. After removing auxiliary categories and inter-category focal loss, ``filler" category's embeddings are widely spread and more overlapped with ``non-filler" category and hard classes. It is demonstrated that our method better models hard classes, which reduces the difficulty of distinguishing the filler words from the non-filler words.

\subsection{Ablation studies}

To verify the effect of our proposed auxiliary categories and inter-category focal loss, we replace the inter-category focal loss by the optimized focal loss used in AF-KWS as a version of ``Ours w/o inter-category focal loss", and remove auxiliary categories as a version of ``Ours w/o auxiliary categories". As Table \ref{tab:ablation} shows, auxiliary categories and inter-category focal loss can effectively improve the performance. 

In order to study the effect of the number of auxiliary categories, we set different number of heads. As shown in Table \ref{tab:hp-head-n}, when the number of auxiliary categories $ h=8 $, the model gets the best performance.

In order to investigate whether the model should pay more attention to samples that misclassify filler words as non-fillers or misclassify non-fillers as fillers, and whether the model should pay more attention to positive or negative samples, we adjust the four weight factors in inter-category focal loss. The results are shown in Table \ref{tab:hp-factor}, the best results are obtained at $ {{\mu _{FN}}}=2 $, $ {{\omega _{FN}}}=2 $, $ {{\mu _{NF}}}=0.5 $, $ {{\omega _{NF}}}=0.5 $, indicating that the model should pay more attention to the error of misclassifying filler words as non-filler words, while errors on positive samples and errors on negative samples can keep the same weight.

\begin{table}[t!]

  \centering
  \begin{tabular}{lccc}
    \toprule
    \multicolumn{1}{c}{\textbf{Method}} & 
    \multicolumn{1}{c}{\textbf{Precision}} & 
    \multicolumn{1}{c}{\textbf{Recall}} & 
    \multicolumn{1}{c}{\textbf{F1}} \\
    \midrule
   Ours & 	                    80.9& 		   81.0& 			81.0  \\
   \quad w/o inter-category focal loss & 	78.3&   77.7& 			78.0 \\
   \quad w/o auxiliary categories & 		80.2& 		   70.4& 			74.9 \\
    \bottomrule
  \end{tabular}
   \caption{Ablation results.}
  \label{tab:ablation}
\end{table}

\begin{table}[t!]

  \centering
  \begin{tabular}{cccc}
    \toprule
    \multicolumn{1}{c}{\textbf{Auxiliary categories $h$}} & 
    \multicolumn{1}{c}{\textbf{Precision}} & 
    \multicolumn{1}{c}{\textbf{Recall}} & 
    \multicolumn{1}{c}{\textbf{F1}} \\
    \midrule
   4    & 	    84.0& 		   76.0&			79.8 \\
   8    & 		80.9& 		   81.0& 			81.0 \\
   12   & 		81.0& 		   80.9& 			80.9 \\
    \bottomrule
  \end{tabular}
   \caption{Filler detection results for the proposed method on the PodcastFillers dataset with different auxiliary categories number.}
  \label{tab:hp-head-n}
\end{table}

\begin{table}[t!]

  \centering
  \begin{tabular}{rrrrccc}
    \toprule
    \multicolumn{1}{c}{\textbf{$ {{\mu _{FN}}} $}} &
    \multicolumn{1}{c}{\textbf{$ {{\omega _{FN}}} $}} &
    \multicolumn{1}{c}{\textbf{$ {{\mu _{NF}}} $}} &
    \multicolumn{1}{c}{\textbf{$ {{\omega _{NF}}} $}} &
    \multicolumn{1}{c}{\textbf{Precision}} & 
    \multicolumn{1}{c}{\textbf{Recall}} & 
    \multicolumn{1}{c}{\textbf{F1}} \\
    \midrule
    2& 		2&	        0.5& 		0.5&    80.9& 	   81.0& 	81.0  \\
    0.5& 	0.5& 	    2& 		    2&	    84.8&	   73.7&    78.9\\
    2& 	    0.5& 		0.5& 		0.5&    80.4&      78.2& 	79.3\\
    0.5& 	  2& 		0.5& 		0.5&    80.0&	   79.0& 	79.5\\
    \bottomrule
  \end{tabular}
   \caption{Filler detection results for the proposed method on the PodcastFillers dataset with different inter-category focal loss factors.}
  \label{tab:hp-factor}
\end{table}

\section{CONCLUSIONS}

In this paper, 
it is verified that 
mixing all other words into one category does not facilitate the detection of filler words. 
We design a filler word detection method that dynamically adds auxiliary categories for training so that the most frequently confused words are explicitly modeled in the network.
For other confusing words left in the ``non-filler" category, we propose an inter-category focal loss which adaptively adjusts the penalty weight between any two categories. 
The ablation experiments prove that the auxiliary categories and the inter-category loss are effective.
Our system achieve an 81.0 F1 score on the PodcastFillers dataset, outperforming other methods by a large margin.

\vfill\pagebreak

\bibliographystyle{IEEEbib}

\bibliography{refs}

\end{document}